\documentclass[aps,superscriptaddress,amsmath,amssymb,twocolumn,showpacs,floatfix,english,noshowpacs,groupedaddress]{revtex4-2}

\usepackage{url}
\usepackage{bm,bbm}
\usepackage{graphicx}
\usepackage{color}
\usepackage[colorlinks=true, urlcolor=blue, linkcolor=blue, citecolor=blue, pdftex]{hyperref}
\usepackage[english]{babel}
\usepackage{soul}
\usepackage{blindtext}
\usepackage{lipsum}

\newcommand{\dagga}{{\phantom{\dagger}}}

\begin{document}

\title{Altermagnetism on the Shastry-Sutherland lattice}

\author{Francesco Ferrari}
\email[]{ferrari@itp.uni-frankfurt.de}
\affiliation{Institute for Theoretical Physics, Goethe University Frankfurt, Max-von-Laue-Stra{\ss}e 1, D-60438 Frankfurt a.M., Germany}
\author{Roser Valent\'\i}
\email[]{valenti@itp.uni-frankfurt.de}
\affiliation{Institute for Theoretical Physics, Goethe University Frankfurt, Max-von-Laue-Stra{\ss}e 1, D-60438 Frankfurt a.M., Germany}

\date{\today}

\begin{abstract}
Motivated by recent developments on altermagnetism, we investigate the Hubbard model on the Shastry-Sutherland lattice, where the onsite repulsion between electrons induces the onset of a staggered magnetic order in which opposite magnetic sublattices are related to each other only by rotations and glide reflections, and not by inversion nor translations. As a consequence, the magnetic phase displays an altermagnetic character, i.e. a finite (non-relativistic) Zeeman splitting of the electronic excitations, despite the absence of a net magnetization. 
By means of a variational Monte Carlo approach based on Jastrow-Slater wave functions, which accounts for electronic correlations beyond the single-particle approximation, we study how $d$-wave altermagnetism shows up in the ground state properties and in the spectral function of the system. We characterize the metal-insulator transition between altermagnetic phases at half-filling and the stability of altermagnetism as a function of doping. The calculation of the single-particle spectral function displays the spin-split nature of the electronic excitations, also within the Mott insulating regime. We discuss possible realizations of the model in the context of
organic $\alpha$- or $\kappa$-(BEDT-TTF)$_2$X charge transfer salts.
\end{abstract}

\maketitle

\section{Introduction}

\begin{figure}[t]
\includegraphics[width=0.55\columnwidth]{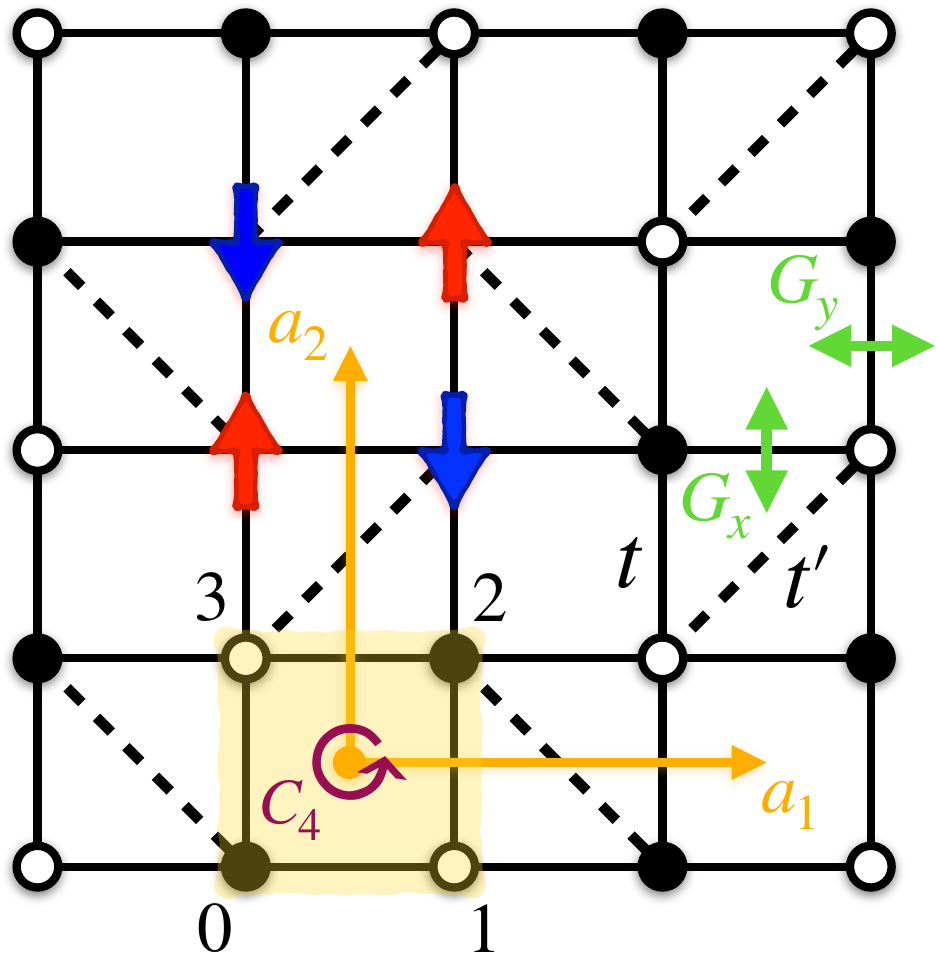}
\caption{\label{fig:lattice} 
Illustration of the Shastry-Sutherland lattice. The yellow box denotes the four sites forming the unit cell ($\eta=0,1,2,3$ sublattices) and the orange arrows represent the Bravais unit vectors $a_1=(1,0)$ and $a_2=(0,1)$. Full/empty dots indicate sites of even/odd sublattices. 
The hoppings of Eq.~\eqref{eq:ham}, $t$ and $t'$, are located on the solid and dashed bonds, respectively. The red/blue arrows depict the $Q=0$ staggered magnetic order. The symmetry operations that connect sublattices of opposite spin-polarization, i.e. $C_4$ rotations and $G_x, G_y$ glide reflections, are marked by the mauve and green arrows.
}
\end{figure}

 Altermagnetism has been recently framed as a new form of magnetism~\cite{vsmejkal2020crystal,vsmejkal2022emerging} which shows no net magnetization, like in antiferromagnets, but fosters a (non-relativistic) Zeeman splitting of the electronic bands, analogously to ferromagnets~\cite{noda2016,hayami2019,yuan2020giant,mazin2021prediction}. This phenomenological distinction is a consequence of the symmetries of the magnetic order and is formalized in the framework of spin groups~\cite{vsmejkal2022emerging}. Concretely, the key aspect to discriminate between antiferromagnetic and altermagnetic (collinear) orders is identifying the set of symmetries which transform sublattices of opposite spin polarizations into each other. In usual antiferromagnets, this is achieved by inversion or translations. Therefore, the combination of such lattice symmetries and time reversal leaves the magnetic order invariant and implies a spin-degeneracy of the electronic bands. This is not the case of altermagnets, where, instead, the effect of time reversal is compensated only by the application of lattice rotations and/or reflections. As a consequence, the altermagnetic character manifests itself in a momentum-dependent spin splitting of the band structure, with an even-parity order parameter with $d$, $g$ or $i$-wave symmetry, which is compatible with the presence of a vanishing net magnetization~\cite{vsmejkal2020crystal,mazin2021prediction,vsmejkal2022beyond,vsmejkal2022emerging,bhowal2024,fernandes2024}.

 Altermagnetism is presently a topic of intensive scrutiny, not only for the possible application of non-relativistic-induced  spin splittings for spintronics purposes~\cite{bai2024altermagnetism} but, on a more fundamental level, for its potential towards unveiling unexplored phases~\cite{zhu2023,chakraborty2024,sim2024}. In this respect, a number of studies have been devoted to the investigation of interacting models where altermagnetism is either imprinted in the local crystallographic symmetries of the model and triggered by the electronic interactions~\cite{maier2023,bose2024,das2024}, or, as recently proposed~\cite{leeb2024}, spontaneously generated by  orbital ordering. While the above studies suggest the realization of interesting altermagnetic phases, most calculations are performed at the single-particle level without including correlation effects (beyond mean-field) that could modify or even prevent the formation of the conjectured order, and some of the proposed models might be somewhat speculative to have them achieved in real materials.
 
In this context, a showcase model in quantum magnetism that has received a large amount of attention in the last decades is the Shastry-Sutherland model~\cite{shastry1981,albrecht1996,miyahara1999,koga2000quantum}  which, in the strong coupling limit, consists of
interacting $S=1/2$ spins on a checkerboard square lattice, where every other square contains an interaction along alternating diagonals (see Fig.~\ref{fig:lattice}); this results in a competition between nearest-neighbor and diagonal exchange interactions, which leads to the formation of distinct ground states depending on the degree of frustration, such as long-range magnetic ordered phases, dimer and plaquette orders, and possibly spin-liquid states~\cite{weihong1999,corboz2013,yang2022,lee2019,viteritti2024}. One of the most studied material realizations of this model is SrCu$_2$(BO$_3$)$_2$~\cite{kageyama1999}, whose behavior under applied magnetic field and pressure illustrates the complex phase diagram of the Shastry-Sutherland model~\cite{miyahara2003theory,zayed2017,lee2019,guo2020quantum,jimenez2021quantum,shi2022discovery}. In view of the richness of phases that the Shastry-Sutherland model entails at the level of the Heisenberg Hamiltonian, a  question that arises is which phases emerge away from the Mott limit of localized electrons~\cite{liu2014}, especially considering that the symmetries of the
underlying lattice could potentially bear altermagnetic order. In this regard, a distorted version of the Shastry-Sutherland lattice is realized by the molecular arrangements of BEDT-TTF molecules in organic charge-transfer salts~\cite{tajima2000,kanoda1997,riedl2022},
such as $\kappa$-(BEDT-TTF)$_2$Cu[N(CN)$_2$]Cl, which displays a momentum dependent spin-splitting of the band structure in its magnetically ordered phase~\cite{naka2019,sumita2023,misawa2023}.

Motivated by these observations, in this work we investigate the ground state and excitation spectra of the fermionic
Hubbard model on the Shastry-Sutherland lattice (Sec.~\ref{subsec:model}). We make use of a variational Monte Carlo approach that allows us to accurately treat the effects of electronic correlations (Secs.~\ref{subsec:vmcgs} and~\ref{subsec:vmcexc}), which have not been considered in previous mean-field studies of altermagnetic models. Starting from a symmetry analysis of the non-interacting band structure (Sec.~\ref{subsec:nonint}), we investigate the onset of altermagnetic order at half-filling as a function of the Hubbard repulsion, uncovering both metallic and insulating altermagnetic phases (Sec.~\ref{subsec:halffilling}). We characterize the altermagnetic order in these two phases and discuss the impact of electronic correlations on the metal-insulator transition. We then assess the stability of altermagnetism as a function of electron and hole doping (Sec.~\ref{subsec:doping}). Finally, we compute the single-particle spectral function, which shows signs of altermagnetism, also in the Mott insulating regime (Sec.~\ref{subsec:spectra}).

\section{Model and Method}
\label{sec:model_and_wf}

\subsection{Hubbard Model} \label{subsec:model}

The focus of this work is on the fermionic Hubbard model~\cite{hubbard1963,arovas2022}
\begin{align}
{\cal H}  = &-t\sum_{\langle i,j \rangle,\sigma} c_{i\sigma}^\dagger c_{j\sigma}^\dagga -t^\prime \sum_{ (i,j)\in d ,\sigma}
c_{i\sigma}^\dagger c_{j\sigma}^\dagga + h.c. \nonumber \\
&+ U\sum_i n_{i\uparrow}n_{i\downarrow}\,
\label{eq:ham}
\end{align}
on the Shastry-Sutherland lattice, whose periodicity can be represented by a square lattice with a four-site unit cell, as depicted in Fig.~\ref{fig:lattice}. The operator $c_{i\sigma}^\dagger$  ($c_{i\sigma}$) creates (annihilates) an electron at site $i$ with spin $\sigma\in\{\uparrow,\downarrow\}$ and $n_{i\sigma}=c_{i\sigma}^\dagger c_{i\sigma}^\dagga$. The Hamiltonian contains two (spin-isotropic) hopping terms, $t$ at first neighbors, and $t'$ residing on a subset $d$ of the square-lattice diagonals. The Hubbard repulsion is denoted by $U$.
The model of Eq.~\eqref{eq:ham} is invariant under the $p4g$ (two-dimensional) space group, which features $C_4$ rotations around the center of the unit cell and reflections with respect to the axes parallel to the diagonal bonds $d$. Additional glide reflections are present, e.g. with respect to the axes of first-neighbor bonds~\cite{lee2019}, as illustrated in Fig.~\ref{fig:lattice}. As an alternative to the above notation, it is sometimes convenient to denote the lattice sites as $(R,\eta)$, i.e. specifying the Bravais vector $R$ of the unit cell and the sublattice kind $\eta=0,1,2,3$. This is particularly useful for the definition of Fourier transforms, e.g. in the case of the annihilation operators ${c_{k,\eta,\sigma}=\frac{1}{\sqrt{N_c}} \sum_R e^{-ikR} c_{R,\eta,\sigma}}$, where $N_c$ denotes the number of unit cells.

In absence of diagonal hoppings ($t'=0$), the Hamiltonian of Eq.~\eqref{eq:ham} reduces to the standard square-lattice Hubbard model, whose ground state at half-filling is insulating and antiferromagnetically ordered  for any $U>0$. The full model with both hopping terms, instead, has been widely investigated in its strong coupling regime ($U\gg t,t'$), where it reduces to the Shastry-Sutherland Heisenberg model~\cite{shastry1981}. Here, tuning the ratio of the resulting exchange couplings, the staggered magnetic order melts due to frustration and leaves room for non-magnetic ground states~\cite{corboz2013,yang2022,lee2019,viteritti2024}, among which an exact product state of singlets on the diagonal bonds is realized~\cite{shastry1981,miyahara1999}. In this work, on the other hand, we explore  the ground state phase diagram of the Hubbard Hamiltonian~\eqref{eq:ham} for small and moderate values of $U/t$, investigating the metal-insulator transition at half-filling and the effects of doping, with a special focus on the altermagnetic properties of the magnetically ordered phase at finite $t'$.
We limit our study to $t'/t\leq 1$ (and $t,t'>0$), such that frustration effects are not strong enough to melt magnetic order in the strong coupling regime~\cite{corboz2013,liu2014}.

\subsection{Variational Monte Carlo: ground state}\label{subsec:vmcgs}

We tackle the Hubbard Hamiltonian of Eq.~\eqref{eq:ham} by means of a variational Monte Carlo approach built on Jastrow-Slater wave functions of the form
\begin{equation}
 |\Psi_0\rangle =  \mathcal{P} \mathcal{J}_{\rm s} \mathcal{J}_{\rm n}  |\Phi\rangle\,,
 \label{eq:psi}
\end{equation}
in which long-range density 
\begin{equation}
\mathcal{J}_{\rm n}=\exp\Big(-\frac{1}{2}\sum_{i,j} v_{i,j} n_i n_j\Big)\label{eq:jas_n}
\end{equation}
and spin
\begin{equation}
\mathcal{J}_{\rm s}=\exp\Big(-\frac{1}{2}\sum_{i,j} u_{i,j} S^z_i S^z_j\Big)
\end{equation}
Jastrow factors are applied to a fermionic uncorrelated state $|\Phi\rangle$, i.e. a Slater determinant~\cite{beccabook}. The variational \textit{Ansatz}~\eqref{eq:psi} provides a flexible and efficient tool to account for electronic correlations, going beyond single-particle approximations (i.e., mean-field methods). In this regard, for instance, the density Jastrow factor $\mathcal{J}_{\rm n}$ can be viewed as a generalization of the Gutzwiller correlator~\cite{gutzwiller1963} and, contrary to the latter, is capable of describing  genuine Mott insulators with finite charge fluctuations~\cite{capello2005,capello2006}. The uncorrelated part of the variational state, $|\Phi\rangle$, can be chosen to encode fundamental ingredients of the physical state, e.g. the presence of symmetry-breaking orders. We take $|\Phi\rangle$ to be the ground state of an auxiliary quadratic Hamiltonian, which contains hopping terms and a staggered magnetic field (with pitch vector $Q=0$)
\begin{align}
\mathcal{H}_{\rm aux} = &-  \sum_{\langle i,j \rangle,\sigma} c_{i\sigma}^\dagger c_{j\sigma}^\dagga 
-\chi^\prime \sum_{ (i,j)\in d ,\sigma}
c_{i\sigma}^\dagger c_{j\sigma}^\dagga + h.c. \nonumber \\
& - h \sum_i  (-)^{\eta_i} (c_{i\uparrow}^\dagger c_{i\uparrow}^\dagga - 
c_{i\downarrow}^\dagger c_{i\downarrow}^\dagga)\,. \label{eq:ham0}
\end{align}
Here, $\eta_i$ denotes the sublattice kind of site $i$ in the convention of Fig.~\ref{fig:lattice}.
The diagonal hopping $\chi'$ and the Zeeman field $h$ are variational parameters. At half-filling, if $h=0$ the ground state of $\mathcal{H}_{\rm aux}$ is a paramagnetic metal  (as long as $\chi'\leq 2$). On the other hand, when $h$ is finite, $|\Phi\rangle$ is an altermagnetic metallic state if $\chi^\prime \leq |h|$ ($\chi^\prime \neq 0$) and an altermagnetic insulator if $\chi^\prime >|h|$.

The calculations are carried out on a lattice of ${N_c=8 \times 8}$ unit cells ($N=256$ sites), with periodic-antiperiodic boundary conditions in the directions of $a_1$ and $a_2$, respectively. The projector $\mathcal{P}$ included in the variational \textit{Ansatz} [Eq.~\eqref{eq:psi}] constrains the wave function to the subspace with total number of electrons $N_e$ and ${S^z_{\rm tot}=\sum_i(n_{i,\uparrow}-n_{i,\downarrow})/2=0}$. Calculations are performed at half-filling, $n=N_e/N=1$, and in presence of finite doping $\delta=n-1$. The variational parameters of the auxiliary Hamiltonian ($\chi'$ and $h$) and the Jastrow factors ($v_{i,j}$ and $u_{i,j}$~\footnote{The Jastrow factors pseudopotentials, $v_{i,j}$ and $u_{i,j}$, are assumed to depend only on the Euclidean distance between sites. For this purpose, we adopt the `non-commensurate' geometry of the Shastry-Sutherland lattice, i.e. the arrangement of magnetic Cu$^{2+}$ ions in SrCu$_2$(BO$_3$)$_2$~\cite{kageyama1999}.}) are numerically optimized by the stochastic reconfiguration method~\cite{sorella2005,beccabook} to obtain the lowest variational energy 
\begin{equation}
 E_0=\frac{\langle \Psi_0|\mathcal{H}|\Psi_0\rangle}{\langle\Psi_0|\Psi_0\rangle}   
\end{equation}
for the Hubbard model~\eqref{eq:ham}.

\begin{figure*}[t]
\includegraphics[width=\textwidth]{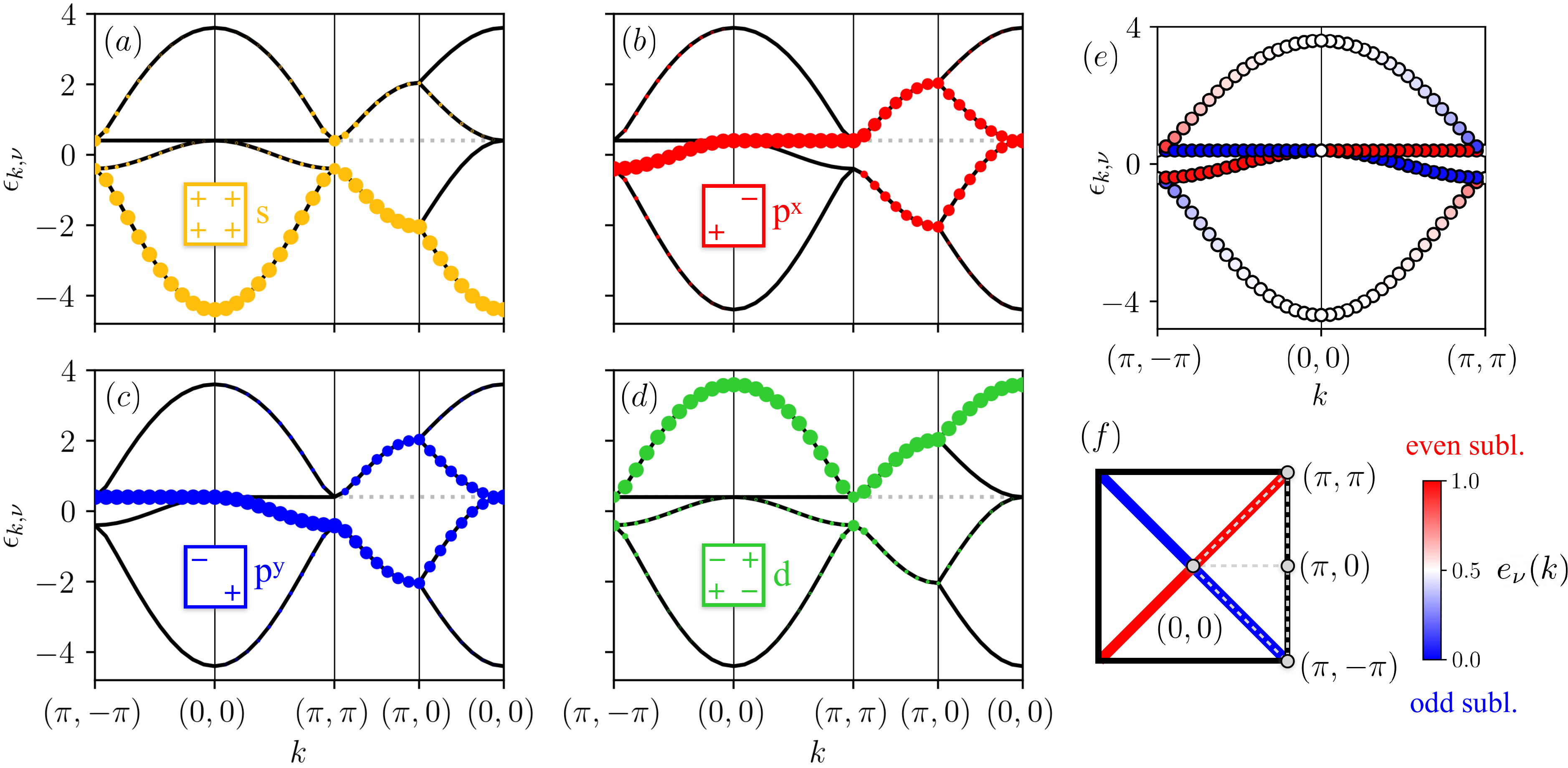}
\caption{\label{fig:bands_para} 
Panels (a-d): band structure of the non-interacting model [Eq.~\eqref{eq:ham} with $U=0$] for $t=1,\ t'=0.4$, plotted along the high symmetry lines of the Brillouin zone, shown in panel (f). 
The size of the dots represents the weights of the Bloch states onto the orbitals of Eqs.~(\ref{eq:swave}-\ref{eq:dwave}), e.g. $|\langle 0 |s^\dagga_k d^\dagger_{k,\nu}| 0 \rangle|^2$ for the case of the $s$-orbital. Projections onto $s$, $p^x$, $p^y$ and $d$ orbitals are displayed in panels (a),(b),(c) and (d), respectively. 
In all panels the Fermi level at half-filling is represented by the horizontal grey dotted line. 
Panel (e): sublattice-resolved band structure. The color of the points corresponds to the value of $e_\nu(k)$ [colorbar in panel (f)], with red and blue colors indicating localization of the Bloch states on even and odd sublattices, respectively. Panel (f): Brillouin zone and sublattice-resolved Fermi surface at half-filling. The Bloch states of the two lines forming the Fermi surface are localized on even/odd sublattice sites.
}
\end{figure*}

\subsection{Variational Monte Carlo: spectral functions}
\label{subsec:vmcexc}

In addition to ground state properties, we employ the method introduced in Ref.~\cite{charlesbois2020} to compute the single-particle spectral function of the Hubbard model~\eqref{eq:ham}. This approach relies on the definition of suitable variational excitations in the sectors with $N_e-1$ and $N_e+1$ electrons, which can be exploited to obtain an approximate expression for the single-particle Green's functions
at momentum $k$, energy $\omega$ and spin $\sigma$:
\begin{equation}
G^{h}_{\eta,\eta'}(k,\sigma,\omega)=\sum_n  \frac{ 
\langle \Psi_0 |c^\dagger_{k,\eta,\sigma} |\Psi^{h}_{k,\sigma,n} \rangle
\langle \Psi^{h}_{k,\sigma,n} |c^\dagga_{k,\eta',\sigma} |\Psi_0 \rangle}{\omega-E_0+E^{h}_{k,\sigma,n}+i\varepsilon}\,, \label{eq:greenh}
\end{equation}
\begin{equation}
G^{e}_{\eta,\eta'}(k,\sigma,\omega)=\sum_n  \frac{
\langle  \Psi_0|c^\dagga_{k,\eta,\sigma} |\Psi^{e}_{k,\sigma,n} \rangle
\langle \Psi^{e}_{k,\sigma,n} |c^\dagger_{k,\eta',\sigma} |\Psi_0 \rangle
}{\omega+E_0-E^{e}_{k,\sigma,n}+i\varepsilon}\,.
\label{eq:greene}
\end{equation}
Here $G^h$ and $G^e$ denote the holes and electrons Green's functions, respectively, which are $4\times 4$ matrices in the sublattice indices $\eta$ and $\eta'$. The sum over $n$ runs over a finite set of variational excitations $\Psi^{h}_{k,\sigma,n}$ ($\Psi^{e}_{k,\sigma,n}$) living in the sector with $N_e-1$ ($N_e+1$) electrons; the respective variational energies are $E^{h}_{k,\sigma,n}$ ($E^{e}_{k,\sigma,n}$). The real parameter $\varepsilon$ introduces a small Lorentzian broadening. 

The approximate excited states entering Eqs.~\eqref{eq:greenh} and~\eqref{eq:greene} are constructed by a Rayleigh-Ritz variational approach in which the Hubbard Hamiltonian is diagonalized within a restricted subspace of variational states~\cite{charlesbois2020}. 
We adopt the basis set proposed in Ref.~\cite{rosenberg2022} and  generalize it to the case of a non-Bravais lattice. For the sake of clarity, let us consider the example of hole excitations. The variational subspace at momentum $k$ and spin $\sigma=\uparrow$ is formed by states of the form
\begin{equation}
\frac{1}{\sqrt{N_c}} \sum_R e^{-ikR} \ T_R c_{0,\eta,\uparrow} B_{0,\eta} |\Psi_0\rangle\,. \label{eq:hole_exc}
\end{equation}
Here, $B_{0,\eta}$ is a certain local operator that conserves the number of electrons and is centered at site $(0,\eta)$; $T_R$ is the translation operator for the Bravais vector $R$. The various states forming the variational basis are then defined by considering different operators $B_{0,\eta}$. The simplest excitations in our basis are the ones in which $B_{0,\eta}=1$, i.e. the bare excitations $c_{k,\eta,\uparrow}|\Psi_0\rangle$, and the ones with $B_{0,\eta}=n_{0,\eta,\downarrow}$. In addition to these, we include a set of states with $B_{0,\eta}=n_{i_{0,\eta},\uparrow} n_{j_{0,\eta},\downarrow}$, where $i_{0,\eta}$ and $j_{0,\eta}$ denote two distinct sites in the neighborhood of site $(0,\eta)$. We restrict them to first-neighbors of $(0,\eta)$ and sites lying across the square lattice diagonals with respect to $(0,\eta)$ (excluding the diagonals perpendicular to the $d$ bonds). The resulting subspace contains $176$ variational states for each momentum and spin. The basis for electron excitations is defined analogously by applying a (uniform) particle-hole transformation to the operators in Eq.~\eqref{eq:hole_exc}. 

The solution of the generalized eigenvalue problem for the Hamiltonian~\eqref{eq:ham} in the variational subspace of hole and electron excitations provides the energies and matrix elements necessary to compute the approximate Green's functions of Eqs.~\eqref{eq:greenh} and~\eqref{eq:greene} (for details see Ref.~\cite{charlesbois2020}). Then, the spectral function can be obtained as
\begin{equation}
A_\sigma(k,\omega)=- \frac{1}{\pi} \sum_\eta {\rm Im}[G^{h}_{\eta,\eta}(k,\sigma,\omega) +G^{e}_{\eta,\eta}(k,\sigma,\omega)]\,, \label{eq:akomega}
\end{equation}
where we have taken the trace over the sublattice indices. The spectral function is thus approximated by a discrete set of delta functions (Lorentzians for finite $\varepsilon$) at each $k$-point. In this respect, the present numerical approach is expected to be less effective for the description of incoherent continua, due to the finiteness of the variational basis. Still, a comparison with exact diagonalization calculations on small clusters showed that the finite set of variational excited states is capable of capturing the main spectral features, also within Mott insulating phases~\cite{charlesbois2020}.

\section{Results}

\subsection{Non-interacting band structure}\label{subsec:nonint}

We begin the discussion by looking at the band structure of the non-interacting model [Eq.~\eqref{eq:ham} with ${U=0}$]. For simplicity, let us drop the spin label at this stage, as the single-particle energies do not depend on $\sigma$. The diagonalization of the non-interacting Hamiltonian yields the bands $\epsilon_{k,\nu}$ shown in Fig.~\ref{fig:bands_para}, with $k=(k_x,k_y)$ and $\nu$ indicating the momentum and the band index, respectively.
On the edges of the Brillouin zone, e.g. along the line $(\pi,\pi)\rightarrow(\pi,0)$, the two upper (lower) bands are degenerate. The third band from the bottom is flat along the $k_x=\pm k_y$ lines, i.e. the diagonals of the Brillouin zone, and displays a quadratic band touching with the second (fourth) band at the zone center (zone corners). For the case $t'=0$ (square lattice limit) the partially-flat bands can be unfolded and correspond to the nested Fermi surface at half-filling. For the remainder of the discussion, we assume that $t'\neq 0$.

It is worth taking a look at the character of the Bloch states. We express their annihilation operators $d_{k,\nu}$ as linear combinations of $c_{k,\eta}$, i.e. $d_{k,\nu}=\sum_\eta \varphi^\nu_{k,\eta} c_{k,\eta}$. We are interested in projecting the various Bloch states at momentum $k$ onto the $s$, $p^x$, $p^y$ and $d$ orbitals formed by combining the different sublattices~\cite{bose2024}
\begin{align}
 &s_k=\frac{1}{2} (c_{k,0}+c_{k,1}+c_{k,2}+c_{k,3})\,, \label{eq:swave}\\
 &p^x_k=\frac{1}{\sqrt{2}}(c_{k,0}-c_{k,2})\,, \label{eq:pxwave}\\
 &p^y_k=\frac{1}{\sqrt{2}}(c_{k,1}-c_{k,3})\,, \label{eq:pywave}\\
 &d_k=\frac{1}{2} (c_{k,0}-c_{k,1}+c_{k,2}-c_{k,3})\,. \label{eq:dwave}
\end{align}
The projected band structure of the non-interacting Hamiltonian is displayed in Fig.~\ref{fig:bands_para}(a-d), with the size of the dots in each panel representing the weights of the Bloch states onto the different orbitals~(\ref{eq:swave}-\ref{eq:dwave}). We note that the lower-most and upper-most bands are mostly localized on the $s$ and $d$ orbitals, respectively. The second and third bands, instead, show contributions also from the $p^x$ and $p^y$ orbitals. Most importantly, the partially flat band along the Brillouin zone diagonals, which is cut by the Fermi level at half-filling, is entirely due to the $p^x$ ($p^y$) orbital on the $k_x=k_y$ ($k_x=-k_y$) line; this fact implies that the corresponding Bloch states are localized on the even (odd) sublattice, as shown also in Fig.~\ref{fig:bands_para}~(e). Here the bands are plotted with a color code that reflects the value of $e_\nu(k)=|\varphi^\nu_{k,0}|^2+|\varphi^\nu_{k,2}|^2$, i.e. the degree of localization of the Bloch eigenvectors on the sites of the \textit{even} sublattice. 

The Fermi surface at half-filling coincides with the two diagonals of the Brillouin zone, $k_x=\pm k_y$, each of them having an opposite sublattice character, as shown in Fig.~\ref{fig:bands_para}(f). This aspect provides a simple argument for the onset of the altermagnetic band splitting in the ordered phase, as discussed in the next section.
We remark that the partially flat band along the diagonal lines is present for any value of $t'/t$. Furthermore, as long as $|t'/t|\leq 2$ the system remains metallic at half-filling and the Fermi surface is always the one shown in Fig.~\ref{fig:bands_para}(f). For $|t'/t|>2$, a gap opens and the system becomes a band insulator.

\subsection{Altermagnetism at half-filling: metallic and insulating phases}\label{subsec:halffilling}

We turn to the discussion of the phase diagram of the Hubbard model Eq.~\eqref{eq:ham} as obtained by variational Monte Carlo. At half-filling ($n=1$), the presence of a finite repulsive interaction $U$ induces the onset of a staggered magnetic order (see Fig.~\ref{fig:lattice}), which is signalled by the stabilization of a finite value of the Zeeman field parameter $h$ upon energy minimization.
In the well-known square lattice limit ($t'=0$), any finite $U$ value is sufficient to induce N\'eel order and turn the system into an insulator~\cite{hirsch1985,white1989,leblanc2015}. In this case, the optimization of the variational parameters yields $\chi'=0$ and $h\neq 0$, i.e. the auxiliary Hamiltonian of Eq.~\eqref{eq:ham0} is the one of an antiferromagnetic insulator. 

\begin{figure}[ht]
\includegraphics[width=\columnwidth]{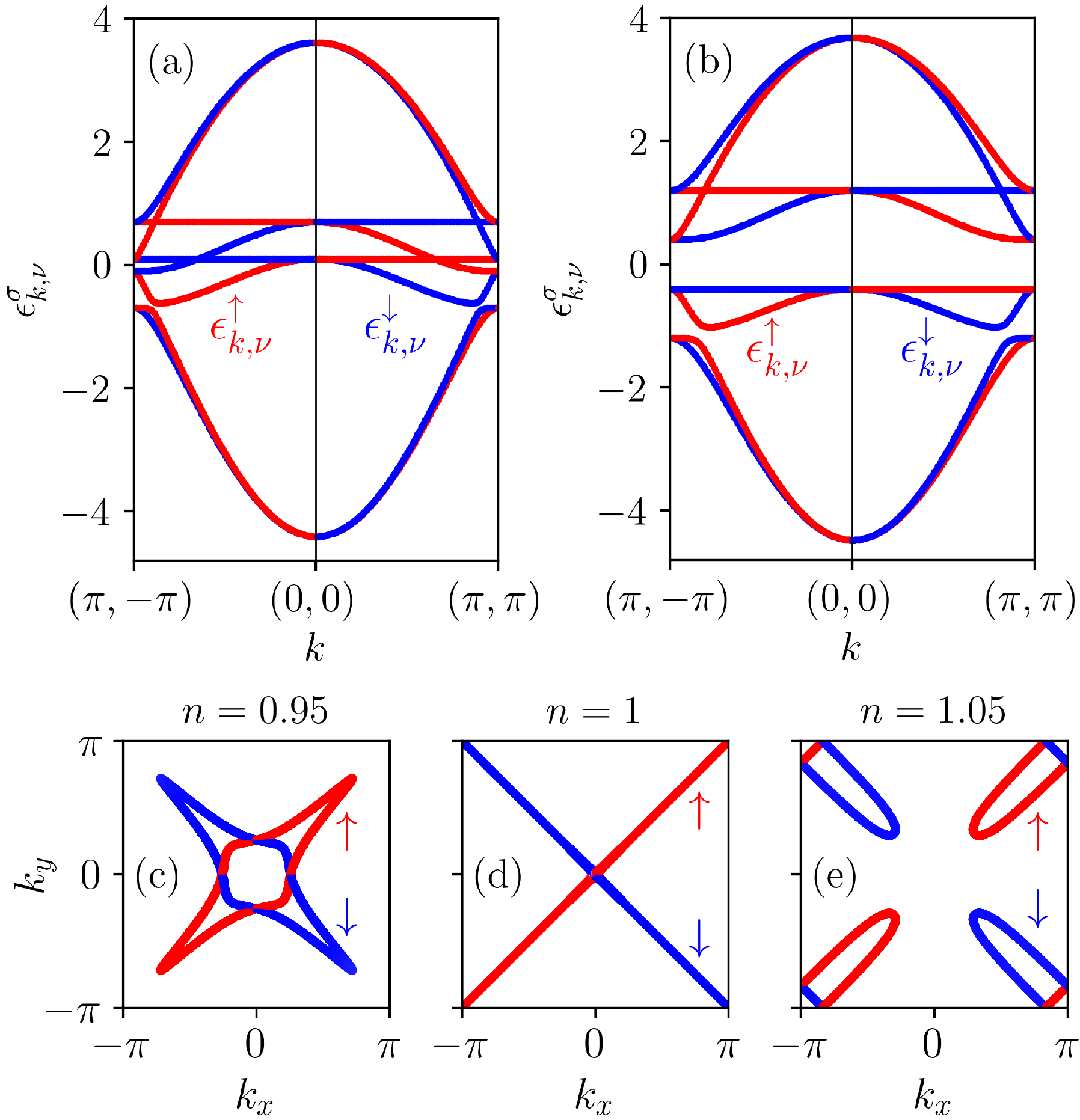}
\caption{Spin-resolved band structure $\epsilon_{k,\nu}^{\sigma}$ of the auxiliary Hamiltonian of Eq.~\eqref{eq:ham0}. Examples for an altermagnetic metal (${\chi'=0.4},{h=0.3}$) and an altermagnetic insulator (${\chi'=0.4},{h=0.8}$) are shown in panels (a) and (b), respectively. The Fermi surface of the metallic case of panel (a) is shown in (c-e), for different fillings. Red and blue colors indicate single-particle energies of $\uparrow$ and $\downarrow$ spins, respectively. \label{fig:bands_magn} 
}
\end{figure}

\begin{figure}[ht]
\includegraphics[width=\columnwidth]{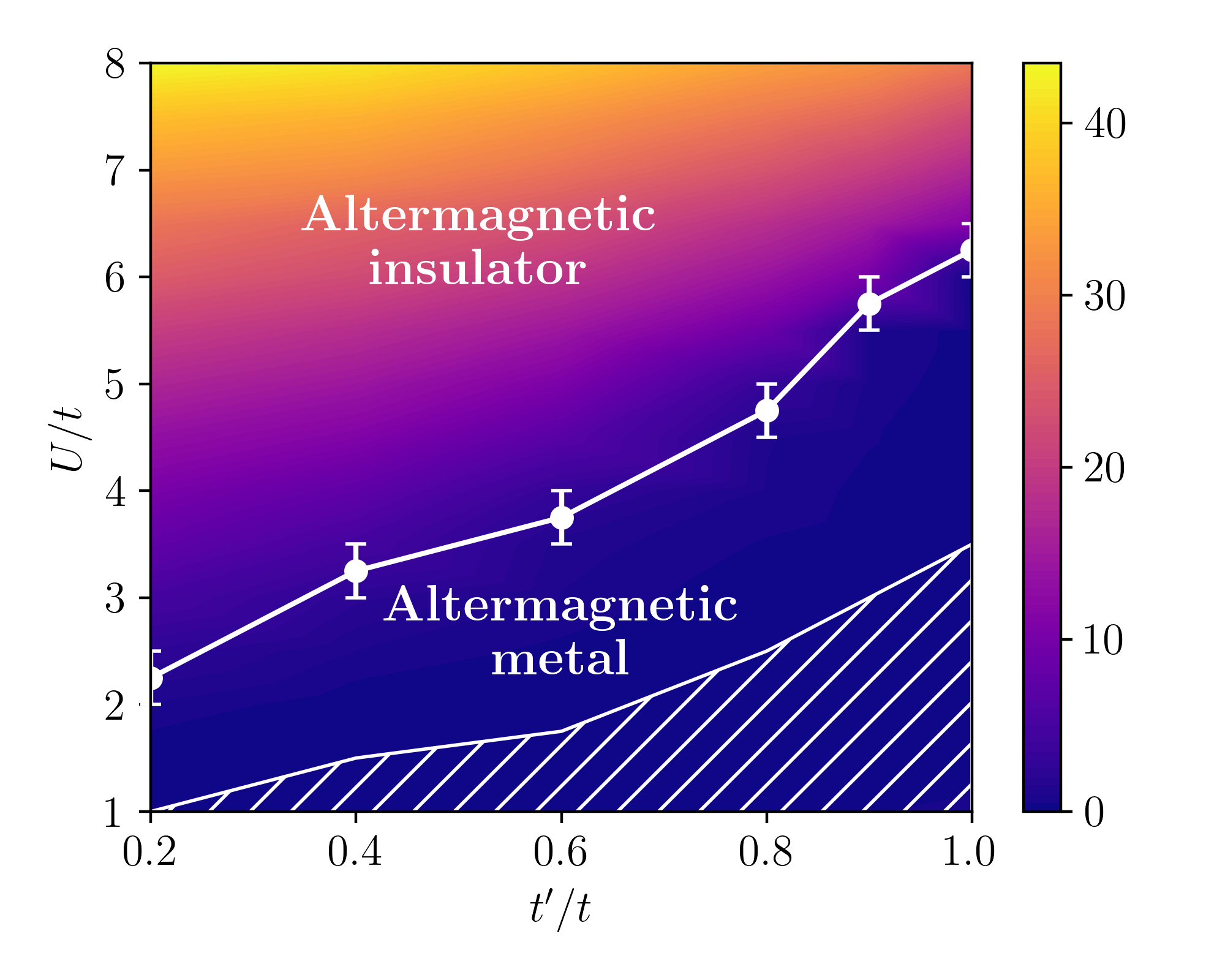}
\caption{
Schematic phase diagram of the Hubbard model~\eqref{eq:ham} at half-filling. The hatched region delimits the metallic phase in which the magnetic order parameter $m$ is very small and hard to distinguish from zero. The colormap shows the value of
$\lim_{q\rightarrow 0} q^2/N(q)$, which is zero (finite) in a metallic (insulating) phase, i.e. can be taken as a proxy for the charge gap of the system~\cite{tocchio2011}. 
\label{fig:phasediag} 
}
\end{figure}


The situation changes when $t'$ is finite. Indeed, the inclusion of the Shastry-Sutherland diagonal hoppings reduces the symmetry of the Hamiltonian from $p4m$ (square lattice) to $p4g$~\cite{parshukov2024}\footnote{The spatial symmetries of the problem are those of the wallpaper group $p4g$ since the Hamiltonian under investigation is strictly two-dimensional. A spin-wallpaper group describing the altermagnetically ordered phases can be defined by supplementing $C_4$ rotations and $G_x,G_y$ glide symmetries with a spin-flip operation (cf. group $\#12.3$ of Ref.~\cite{parshukov2024}).}. Therefore, even and odd sublattices, which correspond to opposite spin orientations of the staggered magnetic order, cease to be connected by translations or inversion. The remaining symmetries which transform the two magnetic sublattices into each other are $C_4$ rotations and the glide reflections with respect to the axes of first-neighbor bonds ($G_x$ and $G_y$ shown in Fig.~\ref{fig:lattice}). Under these conditions, the staggered magnetic order induced by the Hubbard repulsion becomes altermagnetic. This is visible in the band structure of the auxiliary Hamiltonian~\eqref{eq:ham0} where, in addition to a non-zero Zeeman field $h$, a finite $\chi'$ is stabilized: the $\uparrow$ and $\downarrow$ bands split symmetrically in $k$-space, in a $d$-wave fashion, as shown in Fig.~\ref{fig:bands_magn}.
The magnitude of the splitting can be traced back to the degree of localization of the paramagnetic bands on the different sublattices [cf Fig.~\ref{fig:bands_para}, in particular panel (e)]. In this respect, a large splitting is observed for the second and third bands along the $k_x=\pm k_y$ lines, where the non-magnetic Bloch states are entirely localized on the $p^x$ and $p^y$ orbitals. Here, the effect of the Zeeman field is to rigidly shift the bands with opposite sublattice character in opposite directions. On the other hand, no splitting is present for the momenta which are mapped into themselves by rotations and glide symmetries, namely the $k_x=0$ and $k_y=0$ lines and the zone boundaries (i.e., $k_x=\pi$ and $k_y=\pi$ lines). In this case, the combination of the aforementioned point group symmetries with time reversal protects the Kramers' degeneracy of the bands.

\begin{figure}[ht]
\includegraphics[width=\columnwidth]{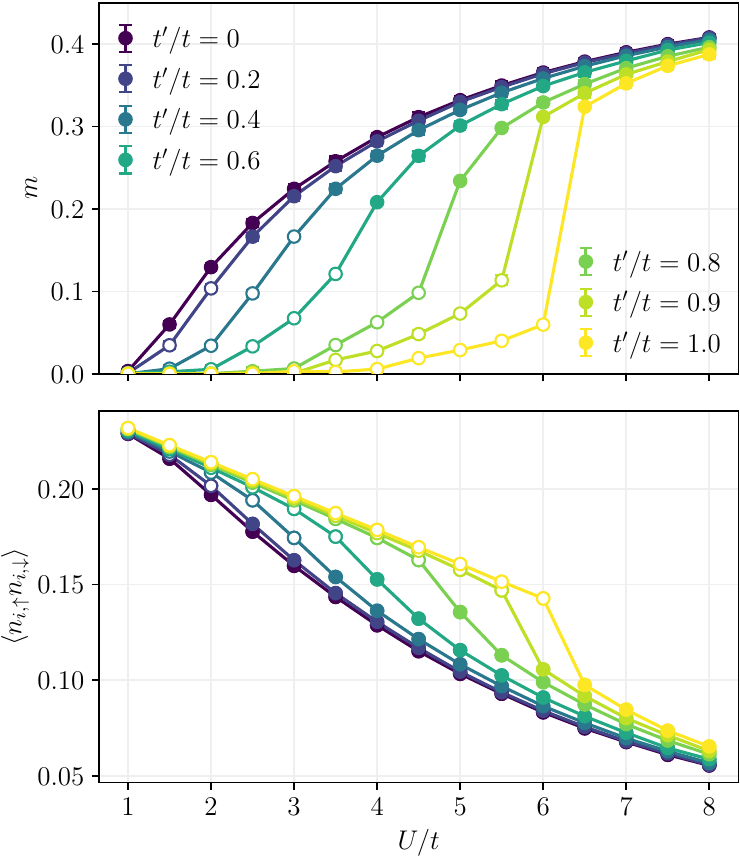}
\caption{Staggered magnetization ($m$, upper panel) and average double occupancy ($\langle n_{i,\uparrow} n_{i,\downarrow}\rangle$, lower panel) as a function of the Hubbard interaction $U/t$ at half-filling. The various sets of data correspond to different values of $t'/t$. Empty (full) dots indicate that the ground state is metallic (insulating).  
\label{fig:magn_double} 
}
\end{figure}

\begin{figure}[ht]
\includegraphics[width=\columnwidth]{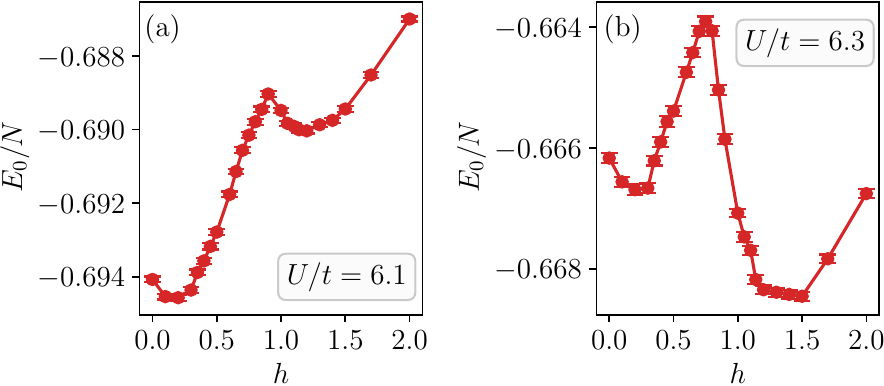}
\caption{Variational energy landscape as a function of the Zeeman field parameter $h$. Results for $t'/t=1$, at $U/t=6.1$ [panel (a)] and $U/t=6.3$ [panel (b)]. \label{fig:landscape} 
}
\end{figure}

As a function of increasing $U/t$, the ground state first becomes an altermagnetic metal and then an altermagnetic insulator, as shown in the phase diagram of Fig.~\ref{fig:phasediag}.  The metallic/insulating nature of the ground state can be directly inferred by looking at the spectrum of the auxiliary Hamiltonian. More quantitatively, one can observe the small-$q$ behavior of the density-density correlation function ${N(q)=\sum_{\eta,\eta'} \langle n_{-q,\eta'} n_{q,\eta} \rangle/4}$, where ${n_{q,\eta}=N_c^{-1/2}\sum_R n_{R,\eta} e^{-iqR}}$ is the Fourier transform of the electron density on the sublattice $\eta$. If $N(q)\sim q$ ($q^2$) for $q\rightarrow 0$, the charge excitations are gapless (gapped)~\cite{feynman1956,tocchio2011}. To characterize the phase transition, we compute the staggered magnetization
\begin{equation}
m=\frac{1}{2N}\sum_i (-1)^{\eta_i} \langle (n_{i,\uparrow}-n_{i,\downarrow})\rangle\,,
\end{equation}
and the average double occupancy $\langle n_{i,\uparrow}n_{i,\downarrow}\rangle$ for different $U/t$ and $t'/t$ values, as shown in Fig.~\ref{fig:magn_double}. Here $\langle\cdot\rangle$ indicates the expectation value over the variational ground state $|\Psi_0\rangle$. 

For small values of the diagonal hopping, i.e. $t'/t \leq 0.6$, the magnetization grows steadily as a function of $U/t$. Also the average double occupancy decreases smoothly, suggesting a continuous metal-insulator transition. While the opening of the charge gap can be reliably located, the critical $U/t$ value necessary to induce magnetic order is harder to estimate. We observe vanishingly small magnetic order at small $U/t$, i.e. a paramagnetic metallic region [hatched area in Fig.~\ref{fig:phasediag}]. However, distinguishing a small $m$ from zero is numerically challenging, and finite-size effects can play a relevant role, as can be seen in the $t'=0$ case, where the value of $m$ seems to vanish at $U/t=1$. Increasing the hopping ratio $t'/t$ induces magnetic frustration in the system and implies a slower growth of $m$ as a function of $U$, which makes the numerical analysis harder. For this reason, the region of very small magnetic order becomes wider for larger values of $t'/t$ and the latter behavior is compatible with the finite-temperature results of Ref.~\cite{liu2014}.

Interestingly, the transition between the altermagnetic metal and the insulator becomes of the first order for larger values of the diagonal hoppings ($t'/t > 0.8$), where we observe a sudden jump in the magnetization and in the average double occupancy. This is further corroborated by the calculation of the energy landscape reported in Fig.~\ref{fig:landscape}, namely the variational ground state energy $E_0$ obtained by optimizing the wave function for different fixed values of the Zeeman field parameter $h$. The energy landscape shows two minima, at small and large $h$ values, which swap their relative positions across the phase transition, showcasing its first-order character. It is worth noting that within a pure mean-field approach the metal-insulator transition is found to be continuous for all values of $t'/t$ considered here. Hence, the Jastrow correlators play a relevant role in determining the first-order behavior at large $t'/t$. These results are thus suggestive of the following scenario: for small $t'/t$, a mean-field-like metal-insulator transition takes place, with a smooth increase of the Zeeman field inducing altermagnetic order in the variational state and gradually leading to a Slater insulating state~\cite{slater1951}, which then crosses over to the Mott regime at large $U$; for large $t'/t$, instead, the slow increase of the magnetic order at small $U/t$ is not sufficient to open a charge gap before the Mott regime kicks in, and thus a first-order transition is observed, with an abrupt increase of $m$ and a jump in the average double occupancy.

\begin{figure}[ht]
\includegraphics[width=\columnwidth]{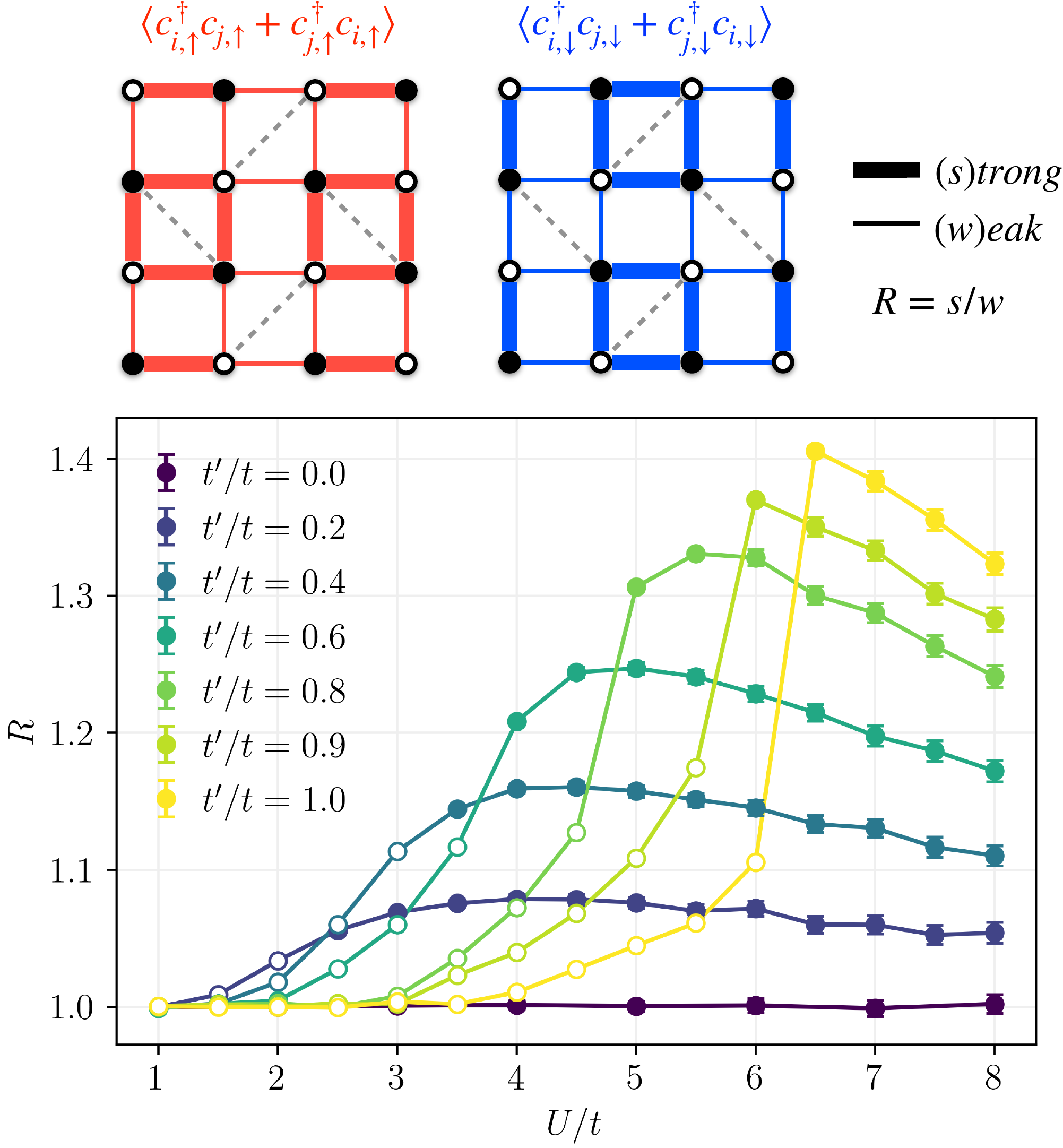}
\caption{Altermagnetic ratio $R$ [Eq.~\eqref{eq:ratio}] as a function of $U/t$, for different values of $t'/t$. Empty (full) dots indicate that the ground state is metallic (insulating). In the upper panel, a sketch of the $\uparrow$ and $\downarrow$ kinetic terms is shown, displaying a splitting between strong and weak contributions. A specular behavior is observed in the two sets of square plaquettes with oppositely oriented diagonals.
$R$ is defined as the ratio between the stronger and the weaker kinetic contributions on each first-neighbor bond.  \label{fig:hopratio} 
}
\end{figure}

To conclude this section, we provide further evidence of the altermagnetic nature of the ordered phases. At the mean-field level, altermagnetism is evidenced by the compensated $k$-dependent splitting of $\uparrow$ and $\downarrow$ bands in the single-particle spectrum. Therefore, one may expect an analogous splitting of the kinetic contributions in the correlated ground state wave function. We thus compute the kinetic terms $\langle c_{i\sigma}^\dagger c_{j\sigma}^\dagga + c_{j\sigma}^\dagger c_{i\sigma}^\dagga \rangle$ at first neighbors, for the optimal variational wave functions at different $t'/t$ and $U/t$ values. The results are summarized in Fig.~\ref{fig:hopratio}. We observe a different behavior in the two sets of square plaquettes that contain a diagonal bond. One set of plaquettes is characterized by strong (weak) kinetic terms in the $\sigma=\uparrow$ ($\downarrow$) channel. The other set of plaquettes, which is connected to the latter by $C_4$ rotations, or $G_x,G_y$ glide reflections, displays the exact opposite behavior. Overall, this gives rise to a $d$-wave-like  spin-polarization of the kinetic contributions in real space (around the center of $C_4$ rotations), analogous to the $d$-wave spin-splitting of the mean-field bands observed in momentum space. We can then take the ratio between the strong and weak kinetic terms on each bond as a real-space marker for altermagnetism, i.e. 
\begin{equation}
 R= \max(r,1/r) \ \ \mbox{ with }
 r=\frac{\langle c_{i\uparrow}^\dagger c_{j\uparrow}^\dagga + c_{j\uparrow}^\dagger c_{i\uparrow}^\dagga \rangle}{\langle c_{i\downarrow}^\dagger c_{j\downarrow}^\dagga + c_{j\downarrow}^\dagger c_{i\downarrow}^\dagga \rangle}\,. \label{eq:ratio}
\end{equation}
Indeed, for $t'=0$, $R=1$ for all $U$ values, i.e. the kinetic terms are isotropic in spin and space, and the system is not altermagnetic, as expected. For finite $t'$, the altermagnetic ratio $R$ first increases and then decreases as a function of $U/t$, denoting the onset of altermagnetism. We again observe a smooth behaviour for small values of $t'/t$, indicative of a continuous transition, with the maximum of $R$ being located in the insulating phase. On the other hand, for large $t'/t$, the ratio jumps when the discontinuous metal-insulator transition is hit.

\subsection{The effect of doping} \label{subsec:doping}

\begin{figure*}[ht]
\includegraphics[width=\textwidth]{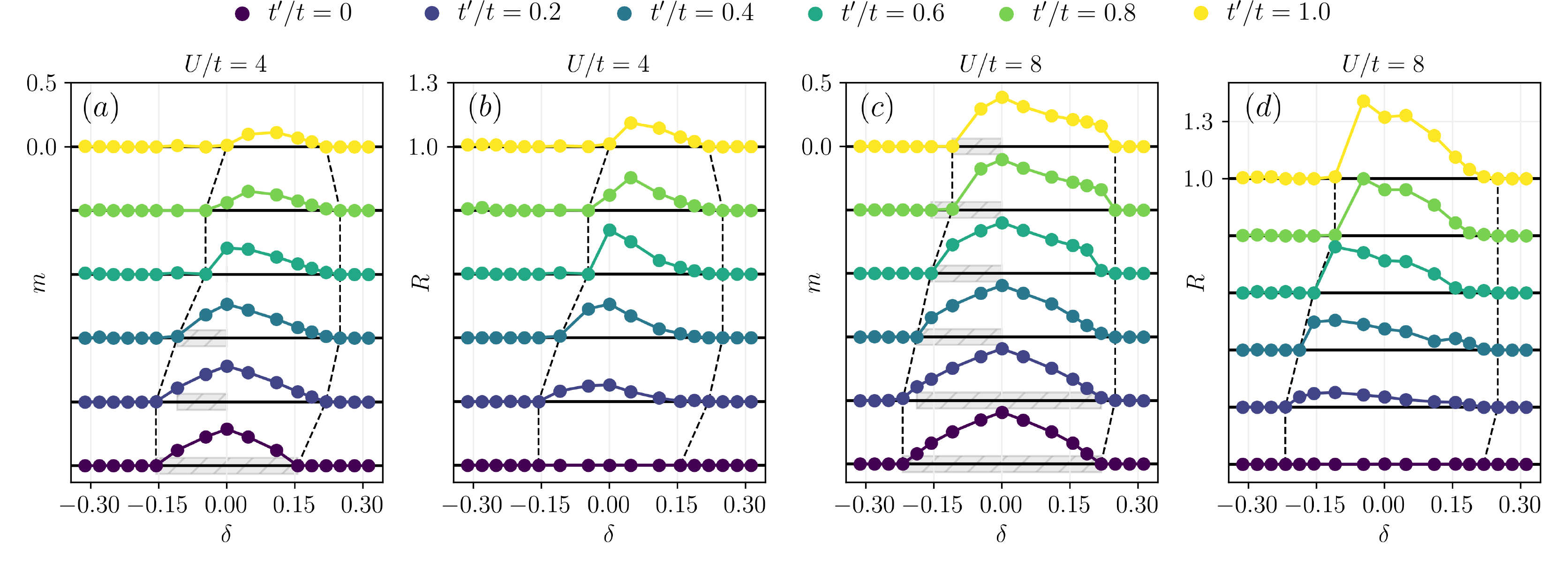}
\caption{Staggered magnetization $m$ and altermagnetic ratio $R$ at $U/t=4$ [panels (a),(b)] and $U/t=8$ [panels (c),(d)], as a function of doping $\delta$. Results for different values of $t'/t$ are shown. Negative (positive) values of $\delta$ refer to hole (electron) doping. The different sets of data of each panel are plotted on the same scale, which is specified in the upper-most plot (i.e., for $t'/t=1$). The solid horizontal lines mark the ground reference values, $m=0$ in panels (a),(c) and $R=1$ in panel (b),(d).
The dashed lines delimit the region with finite magnetic order. The regimes in which system is prone to phase separation are indicated by the hatched grey areas in panels (a),(c).\label{fig:doping} 
}
\end{figure*}

We assess the stability of the altermagnetic phases when the system is doped away from half-filling. We consider the cases with $U/t=4$ and $U/t=8$, and track the evolution of the altermagnetic order as a function of $\delta=n-1$. The results are summarized in Fig.~\ref{fig:doping}, where the staggered magnetization $m$ and the altermagnetic ratio $R$ are plotted as a function of doping. In our notation, $\delta<0$ ($>0$) indicates hole (electron) doping. We note that for $t'=0$ the system reduces to the standard Hubbard model on the square lattice and it is thus particle-hole symmetric~\cite{arovas2022}. For finite $t'$, instead, the staggered particle-hole transformation, $c_{i,\sigma} \mapsto (-)^{\eta_i}c^\dagger_{i,\sigma}$, maps the model at doping $\delta$ to the model at doping $-\delta$, but changes $t' \mapsto -t'$. Thus the system ceases to be particle-hole symmetric, as clearly shown in the results of Fig.~\ref{fig:doping}. 

Let us first discuss the results at large $U$, i.e. $U/t=8$ [Fig.~\ref{fig:doping}(c,d)], where the system is in the altermagnetic insulating phase at half-filling for all $t'/t$ values considered here. Away from half-filling, the system immediately becomes metallic and the staggered magnetization $m$ starts decreasing as a function of doping, see Fig.~\ref{fig:doping}(c). The region of stability of the altermagnetic order to hole doping shrinks when $t'/t$ is increased, while in the case of electron doping it remains almost constant. For hole doping, the altermagnetic ratio $R$, shown in Fig.~\ref{fig:doping}(d), first increases and then decreases, reaching $R=1$ when $m$ vanishes. The disappearence of the altermagnetic order with hole doping is smooth for small $t'/t$ and abrupt for large $t'/t$. In presence of electron doping, we observe a similar behavior for the magnetization $m$, which jumps to zero for large $t'/t$. On the other hand, the altermagnetic ratio displays a smooth behavior for all $t'/t$ values.

It is important to note that our variational results indicate that the altermagnetic metal at finite doping is unstable to phase separation. Considering as an example the case of hole doping, phase separation can be detected by inspecting the behavior of the energy per hole, $e(\delta)=[E_0(\delta)-E_0(\delta=0)]/|\delta|$, as a function of $\delta$ \cite{emery1990,tocchio2016}. For a stable phase, $e(\delta)$ is monotonic, while in presence of phase separation (in a finite-sized system) it displays a minimum at a certain value of $\delta$. This indicates that the system can gain energy by forming hole-rich and hole-poor regions. At $U/t=8$, we find  that large regions of phase separation in the presence of hole doping, marked by the grey hatched areas in Fig.~\ref{fig:doping}(c). On the contrary, for electron doping phase separation is found only for $t'/t \leq 0.2$. We point out, however, that variational approaches like the one employed in this work tend to overestimate the tendency to phase separate, as the accuracy of the variational \textit{Ansatz} can differ for different electron densities $n$, thus affecting the above analysis based on the calculation of $e(\delta)$. In this respect, for instance, the authors of Ref.~\cite{tocchio2016} showed that in the square lattice limit ($t'=0$), for relatively large $U$ values (e.g., $U/t=8$ considered here) the region of phase separation shrinks considerably when improving Jastrow-Slater variational results by performing more accurate fixed-node Monte Carlo calculations. 

We now turn to the results at smaller $U$ values, i.e. $U/t=4$, shown in Fig.~\ref{fig:doping}(a,b).
In this case, the system is insulating at half-filling for $t'/t\leq 0.6$ and metallic for for larger values of the diagonal hopping, as reported in Fig.~\ref{fig:magn_double}. In presence of finite doping, the insulating phase at $t'/t\leq 0.6$ immediately turns to a metallic altermagnet. Then the altermagnetic order decreases both for hole and electron dopings, displaying a larger stability in the latter case.  An interesting effect is observed when doping the altermagnetic metal found at half-filling for $t'/t=0.8$ and $t'/t=1$. In case of electron doping, the altermagnetic order is enhanced, especially in the case $t'/t=1$, where the magnetization $m$ is vanishingly small at half-filling and clearly becomes finite for $\delta>0$. Both $m$ and $R$ first increase and then smoothly decrease to zero as a function of $\delta$. Finally, we note that phase separation at $U/t=4$ is found only at small $t'/t$ and, except for the case $t'=0$, only in the case of hole doping.

\begin{figure*}[ht]
\includegraphics[width=\textwidth]{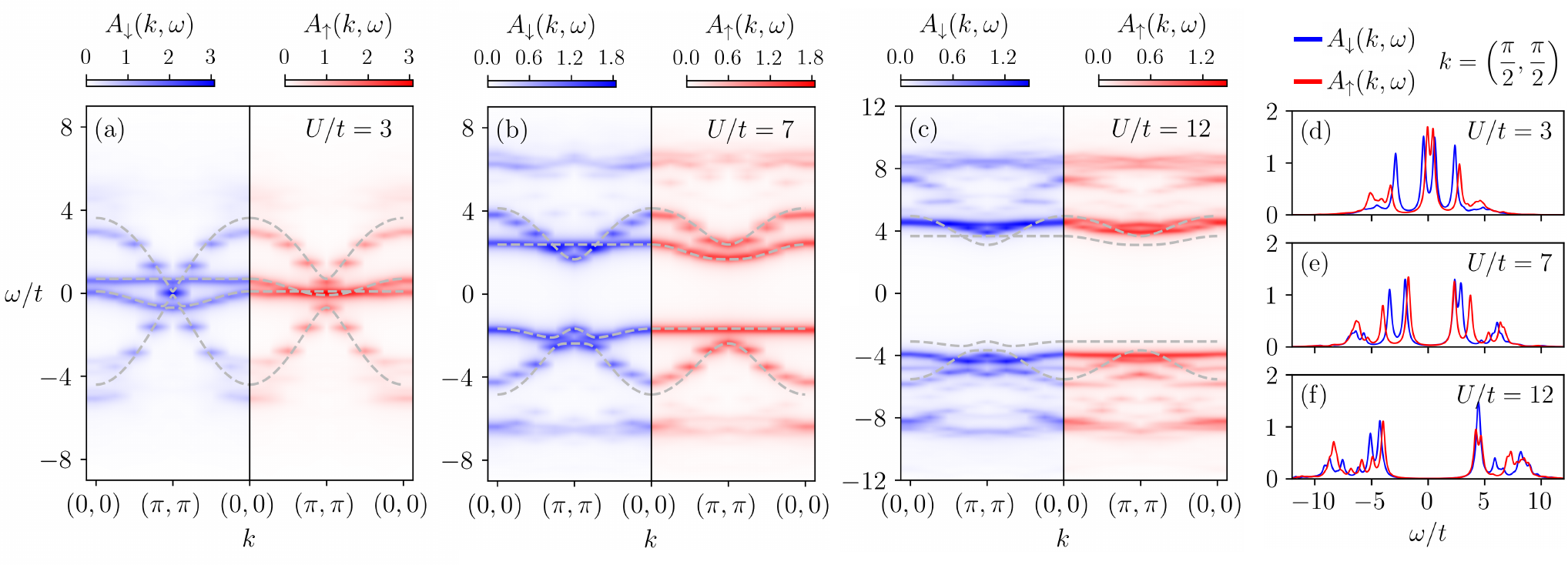}
\caption{\label{fig:spectra_hf} Spectral function $A_\sigma(k,\omega)$ at half-filling ($n=1$) for $t'/t=0.4$. Results for the altermagnetic metal [$U/t=3$, panel (a)] and the altermagnetic insulator [$U/t=7$ and $12$, panels (b) and (c)] are shown. The panels are symmetrically split to show the $\sigma=\downarrow$ (left side) and $\sigma=\uparrow$ (right side) contributions along the same $k$-path, with blue and red color schemes, respectively.
The dashed lines represent the spin-resolved bands of the optimal auxiliary Hamiltonian [Eq.~\eqref{eq:ham0}]. A
Lorentzian broadening $\varepsilon=0.2t$ is used. Panels (d-f) show $A_\downarrow(k,\omega)$ and $A_\uparrow(k,\omega)$ at $k=(\pi/2,\pi/2)$, for the same $U/t$ values of panels (a-c), respectively. The Fermi level is at $\omega=0$.}
\end{figure*}

\subsection{Spectral functions}
\label{subsec:spectra}

We conclude the results section by presenting the spectral function of the Hubbard model as obtained by variational Monte Carlo, as it provides the most direct evidence of altermagnetism~\footnote{For the calculations of the spectral function we use periodic boundary conditions both in the $a_1$ and $a_2$ directions, in order to preserve the point group symmetry of the Shastry-Sutherland lattice. In the case $t'/t=0.4$ and $U/t=3$, the system is metallic at half-filling and the spectrum of $\mathcal{H}_{\rm aux}$ is open-shell. For computational reasons, we split the degeneracy at the Fermi level by adding a tiny hopping of $10^{-6}$ across the diagonals of the empty squares (in $\mathcal{H}_{\rm aux}$). This has no visible effects on the physical results.}. Indeed, in the altermagnetic phases a splitting of the $A_\uparrow(k,\omega)$ and 
$A_\downarrow(k,\omega)$ components along certain $k$-directions is observed, analogously to the spin-split band structure of the auxiliary Hamiltonian of Eq.~\eqref{eq:ham0} (see Fig.~\ref{fig:bands_magn}). The symmetry of the altermagnetic order on the Shastry-Sutherland lattice implies that the spectral features for $\sigma=\uparrow$ and $\downarrow$ components are related to each other by, e.g., a $C_4$ rotation in $k$-space.

\begin{figure}[ht]
\includegraphics[width=\columnwidth]{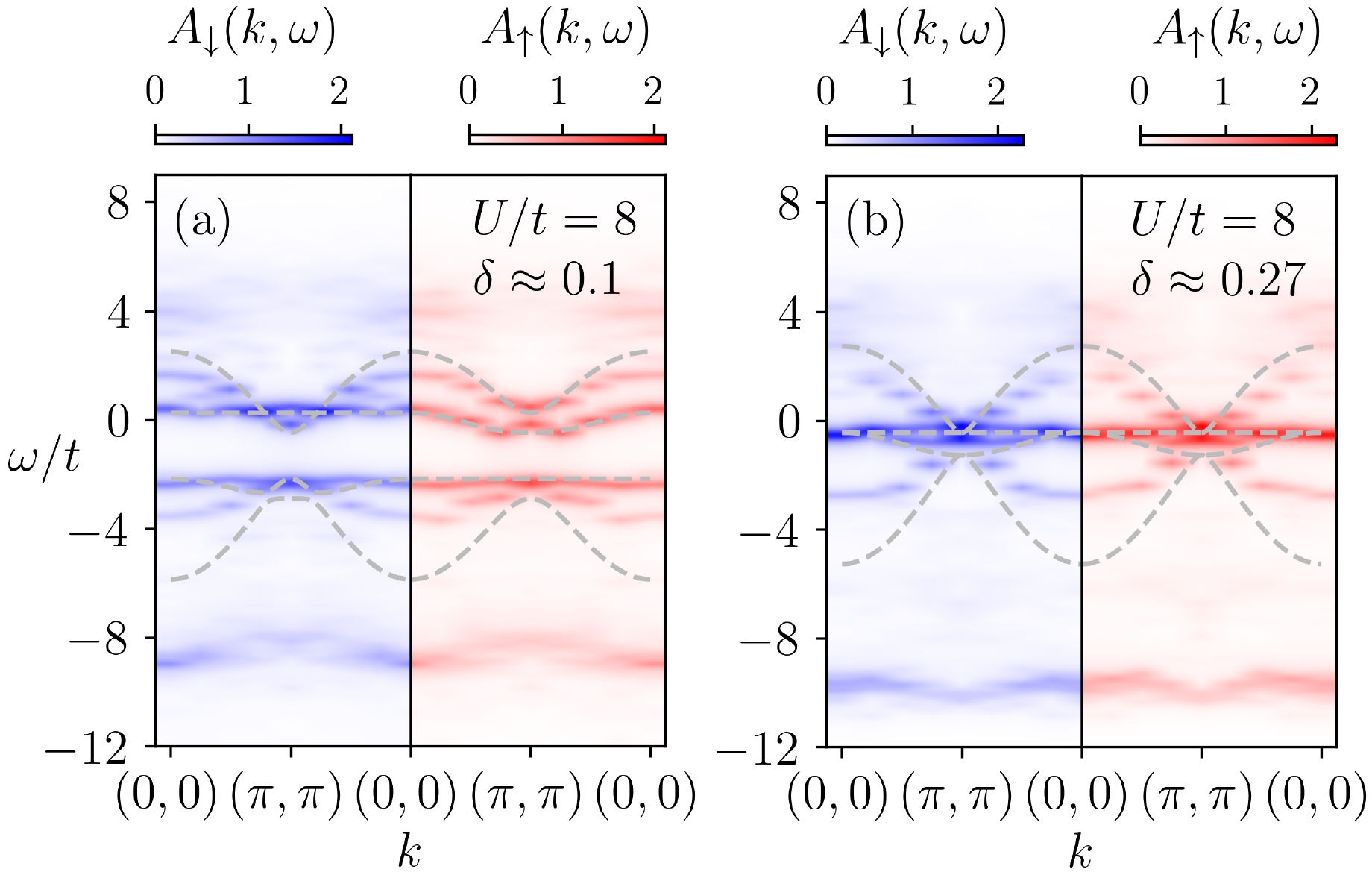}
\caption{The same as in Fig.~\ref{fig:spectra_hf}, for $t'/t=0.4$, $U/t=8$ and two different fillings in the electron doped case. The results of panel (a) are obtained in the altermagnetic metallic phase at small doping ($N_e=282$, $\delta\approx0.1$); the results in panel (b) are in the paramagnetic metallic phase at large doping ($N_e=326$, $\delta\approx0.27$). \label{fig:spectra_doped}}
\end{figure}

In Fig.~\ref{fig:spectra_hf}, we show the spectral function at half-filling for $t'/t=0.4$ and different values of $U/t$. The spin-resolved $A_\uparrow(k,\omega)$ and 
$A_\downarrow(k,\omega)$ are plotted along the $k_x=k_y$ diagonal of the Brillouin zone. In the metallic regime at $U/t=3$, the spectrum closely resembles the one of the auxiliary Hamiltonian, with well-defined quasiparticle bands, especially close to the Fermi level ($\omega=0$). The altermagnetic splitting is showcased by, e.g., the second and third bands, one dispersive and one flat, which are pushed in opposite energy directions in the $\uparrow$ and $\downarrow$ channels by the presence of finite altermagnetic order.
Increasing $U/t$ and entering the insulating regime, we observe the gap opening and a progressive broadening of the spectral features into the upper and lower Hubbard bands. The altermagnetic splitting is visible also within the Mott insulating phase, where the excitations are incoherent, e.g. for $U/t=12$. In Fig~\ref{fig:spectra_hf}(d-f), we report a comparison of $\uparrow$ and $\downarrow$ spectral functions at fixed momentum $k=(\pi/2,\pi/2)$ which helps visualizing the  splitting for all $U/t$ values.

Finally, in Fig.~\ref{fig:spectra_doped}, spectral functions at finite electron doping are presented. For small $\delta$, the system is in the altermagnetic metallic phase and the spectral function displays the splitting between the spin channels [see Fig.~\ref{fig:spectra_doped}(a)].
While close to the Fermi energy the dispersion is well-captured by the band structure of the auxiliary Hamiltonian, away from $\omega=0$ the signal becomes incoherent and the altermagnetic splitting is seemingly reduced. On the other hand, at large electron dopings altermagnetism is suppressed. As shown in Fig.~\ref{fig:spectra_doped}(b), the $A_\uparrow(k,\omega)$ and 
$A_\downarrow(k,\omega)$ spectral functions are  symmetric and no spin-splitting is observed.

\section{Conclusions}

In this work, we investigated the properties of the Hubbard model on the Shastry-Sutherland lattice in the regime of moderate frustration, as it provides a minimal setting to study altermagnetism in a correlated system. We made use of a variational Monte Carlo approach, based on Jastrow-Slater wave functions, to investigate ground state (and spectral) properties beyond mean-field-like approximations. 

At half-filling, when the Shastry-Sutherland diagonal hopping $t'$ is finite (i.e., away from the square-lattice limit), the system is found to be first metallic for a finite range of $U/t$, and then insulating for larger values of the Hubbard interaction. Most importantly, the onsite repulsion induces the onset of a staggered magnetic order, which bears altermagnetic properties due to the symmetry of the underlying lattice. We studied the metal-insulator transition between altermagnetic phases at half-filling, as a function of $U/t$. While for small values of the diagonal hoppings, the transition is found to be continuous and Slater-like, for sufficiently large $t'/t$ a first-order jump of the magnetization is detected in correspondence of the gap opening. The change in the nature of the transition testifies the importance of electronic correlations, as it is not captured by a standard mean-field approach. In addition to the magnetic order parameter, we provided a static measure of altermagnetism by computing the difference of the first-neighbors kinetic terms in the two spin channels (parallel to the magnetic order direction). We investigated also the effects of doping, assessing the stability of the altermagnetic order away from half-filling. Finally, we computed the spin-resolved spectral function, providing evidence of the altermagnetic splitting of the spectral features, both in the metallic ground state with well-defined quasiparticle excitations, and in the highly-correlated Mott regime.

These results present a new avenue to explore correlation-induced altermagnetic phases. Possible material candidates where such phases could be realized are, for instance, $\alpha$- or $\kappa$-(BEDT-TTF)$_2$X~\cite{tajima2000,kanoda1997,riedl2022} charge-transfer salts, whose underlying lattice structure can give rise to a distorted version of the Shastry-Sutherland lattice. Although $\kappa$-(BEDT-TTF)$_2$X salts are actually closer to the regime of strong dimer hopping (i.e., $t' \gg t$) and away from half-filling~\cite{koretsune2014,guterding2016}, the versatility of these materials as a function of pressure, strain and doping may open up the possibility to tune their properties towards altermagnetic phases related to the ones observed in this work. On the theoretical side, instead, the possible presence of superconductivity in the doped regime warrants further investigation, also in view of the possibility of realizing pair-density wave states~\cite{sim2024,chakraborty2024,sumita2023}. A more accurate variational approach including backflow correlations could be suitable for this purpose~\cite{tocchio2008}.

\section*{Acknowledgments}
We would like to thank F. Becca, J. Knolle, I. I. Mazin, J. Profe, J. Sinova,  L. \v{S}mejkal and A. Valadkhani for useful discussions, and P. P. Stavropoulos for providing mean-field results for comparison.
We gratefully acknowledge support by the Deutsche Forschungsgemeinschaft (DFG, German Research Foundation) for funding through TRR 288—422213477 (project B05).

\end{document}